# Long-Term Variations in Sunspot Characteristics

A. G. Tlatov

*Kislovodsk Mountain Astronomical Station, Central (Pulkovo) Astronomical Observatory,
Russian Academy of Sciences, Kislovodsk, Russia*
e-mail: tlatov@mail.ru

Relative variations in the number of sunspots and sunspot groups in activity cycles have been analyzed based on data from the Kislovodsk Mountain Astronomical Station and international indices. The following regularities have been established: (1) The relative fraction of small sunspots decreases linearly and that of large sunspots increase with increasing activity cycle amplitude. (2) The variation in the average number of sunspots in one group has a trend, and this number decreased from ~12 in cycle 19 to ~7.5 in cycle 24. (3) The ratio of the sunspot index (Ri) to the sunspot group number index ($G_{gr}$) varies with a period of about 100 years. (4) An analysis of the sunspot group number index ($G_{gr}$) from 1610 indicates that the Gnevyshev–Ohl rule reverses at the minimums of secular activity cycles. (5) The ratio of the total sunspot area to the umbra area shows a long-term variation with a period about eight cycles and minimum in cycles № 16-17. (6) It has been indicated that the magnetic field intensity and sunspot area in the current cycle are related to the amplitude of the next activity cycle.

## 1. INTRODUCTION

The Gnevyshev–Ohl (G–O) rule, which was valid for about 150 years (beginning from cycle 10), was violated in cycles 22 and 23. This can indicate that the solar cyclicity regime will change, which possibly took place previously (Vitinsky et al., 1986). The G–O rule is violated during the decline stage of the secular cycle and can indicate that activity will decrease during a long period similar to the Maunder minimum. It is still unknown why prolonged activity cycles exist. The characteristic periods of these cycles (Hathaway, 2010) and variations in sunspot characteristics (magnetic fields, area, group properties, etc.) during secular cycles are still among the problems to be solved.

The aim of this work is to trace relative variations in the properties of sunspots and sunspot groups with different areas in solar activity cycles and in the secular activity cycle.

## 2. VARIATION IN THE RELATIVE CONTRIBUTION OF DIFFERENT SUNSPOTS TO ACTIVITY INDICES

As initial data for this analysis, we took daily observations of sunspot groups at Kislovodsk Mountain Astronomical Station (GAS) from 1954 to 2012 and other data. In addition to the coordinates and area, the number of umbrages and pores ($N_{sp}$), participating in the calculation of the Wolf number, as well as the area of the maximal sunspot in a group ($S_{max}$), are also present in the GAS data. This makes it possible to analyze different activity indices depending on the group or maximal sunspot area. An analysis of the total number of small and large groups indicates that small and large sunspots differently contribute to the Wolf number. The relative number of small sunspots decreases, depending on the activity cycle amplitude, and the fraction of large sunspots increases with increasing activity cycle amplitude. This conclusion does not confirm the conclusion drawn in (Lefe'vre and Clette, 2011) that small sun spots were rarely encountered in cycle 23.

The relative contribution of large sunspot groups ($S > 500$ millionths of solar hemisphere, msh) to the Wolf number increases with increasing activity cycle amplitude $W_{max}$: $W_{500}/W_{tot} = 0,065 + 0,001 \cdot W_{max}$, $R = 0,93$. The contribution of small sunspot groups ($S < 50$ msh) decreases with increasing activity cycle amplitude: $W_{50}/W_{tot} = 0.35 - 6.9 \cdot 10^{-4} \cdot W_{max}$, $R = -0,74$.. Such a regularity is also valid for the total number of groups, including the maximal sunspots with areas of $S_{max} < 20$ msh (small sunspots) and $S_{max} > 700$ msh (large sunspots) (Fig. 1). The relative number of small and large sunspot groups differs, depending on the increase in the activity cycle amplitude.

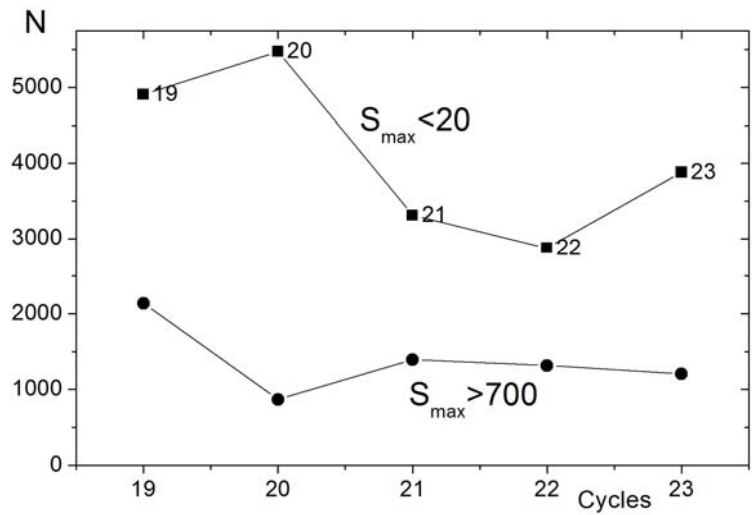

**Fig. 1.** Number of small and large sunspots according to the Kislovodsk data.

## 3. LONG-TERM VARIATIONS IN THE SUNSPOT NUMBER IN GROUPS

The average number of umbrages and pores in sunspot groups decreased monotonically from cycle to cycle during the last five activity cycles (Fig. 2). This is especially pronounced for medium and large sunspots ($S > 50$ msh). Umbras and pores usually play the main role in the calculation of the $W$ index for medium and large sunspots. The average area of individual umbra possibly increased during this period, and their number decreased in this case.

These variations can be verified based on other activity indices. As is known, the number of groups with factor 10 and the total sunspot (umbra) number are taken into account when the Wolf number is calculated. At the same time, the index of the sunspot group number exists (Hoyt and Schatten, 1998).

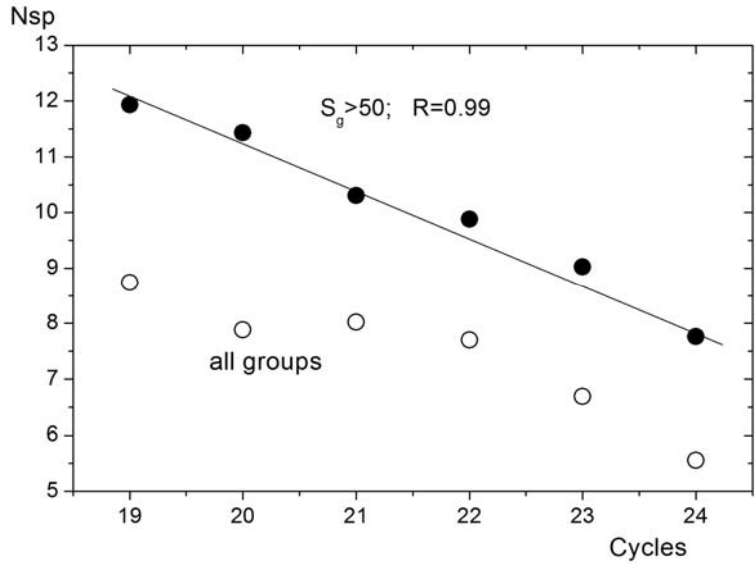

**Fig. 2.** Variation in the average sunspot number in a group during a cycle.

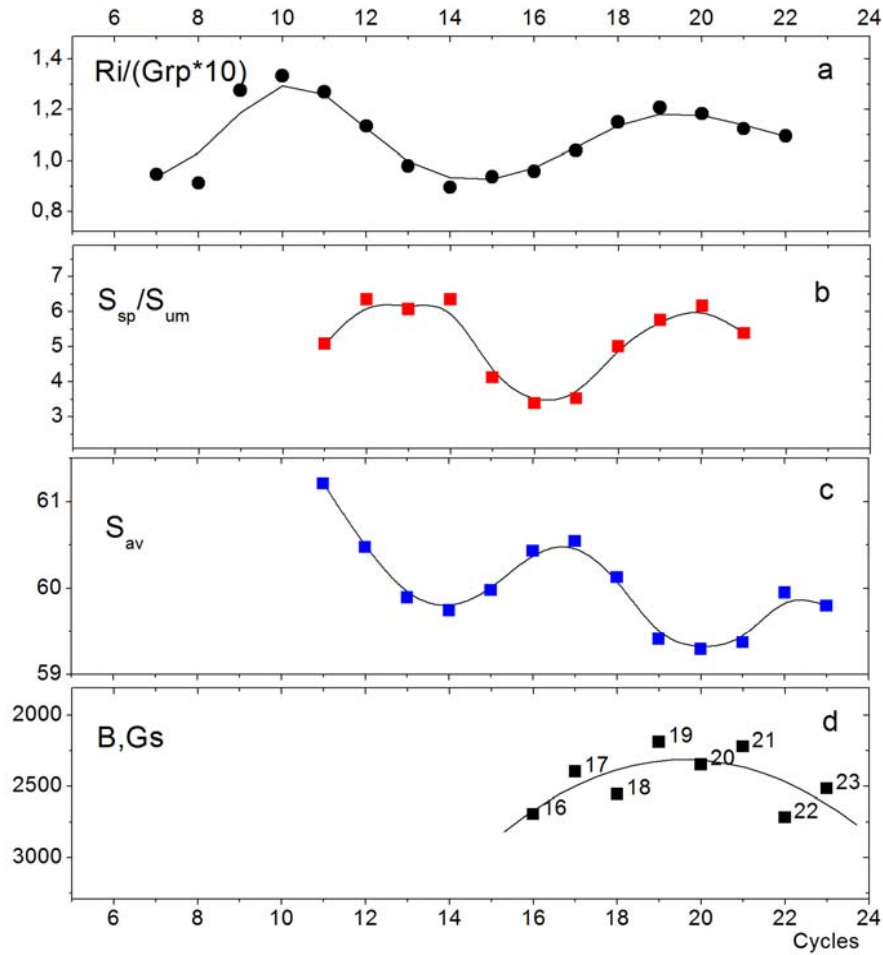

**Fig. 3.** (a) Ratio of the Wolf number to the group number index during cycles. (b) The ratio of the group area to the umbra area according to the RGO data. (c) Variations in the sunspot group area in the range *S*: 30–100 msh (average for a cycle), according to the RGO (cycles 11–20) and GAS (cycles 21–23) data. (d) Variations in the average magnetic field strength according to the MNTW observatory data for sunspots with areas *S* > 100 mhs from 1915 to 2002.

Figure 3a presents the variation in this ratio from 1748, found from daily data and averaged over the solar cycles. There exists a long-term variation with a period of about ten solar cycles. The number of sunspots in one group was maximal in cycles 10 and 19. Based on the sunspot group characteristics at the Greenwich observatory (RGO) (http://solarscience.msfc.nasa.gov), we can reconstruct the ratio of the total sunspot area to the umbra area (Fig. 3b). This ratio also shows a long-term variation with a slightly shorter period (about eight cycles); however, the maximum falls on cycle 13 and 20 and minimum falls on cycle 16-17. Variations also exist in the relative contribution of the areas of different groups, which is confirmed by the variations in the sunspot group area ranging from 30 to 100 msh on average over the cycle. Variations with close periods also exist for other ranges of areas.

Figures 3a–3c indirectly confirm the conclusion that the number of sunspots in one group in cycle 19 is large (Fig. 2). However, the average magnetic field strength in umbrages would decrease in this case since the magnetic field of umbra increases with increasing their area (Vitinsky et al., 1986).

Such a tendency is observed in Fig. 3d for the average magnetic field strength in the cycle according to the Mount Wilson (MNTW) observatory data. The magnetic field ($B$) increased from the middle of the past century to cycle 22. In this case, the magnetic field strength in even cycles was on average higher than in odd cycles.

The long-term variations in the sunspot magnetic field are not random and are directly related to the activity cycle amplitude. The relationship between the magnetic field and area was found for cycles № 15–19 from the MNTW data: $B = a + b\log(S)$ (Petsov et al., 2013), where $a$ and $b$ are the coefficients varying from cycle to cycle. Figure 4 presents the relationship between the product of the total area logarithm during a cycle into coefficient $b$: $b\log(\sum S)$ in a given cycle and the amplitude of the ext activity cycle ($W_{n+1}$). The correlation between these parameters is high ($R = 0.96$).

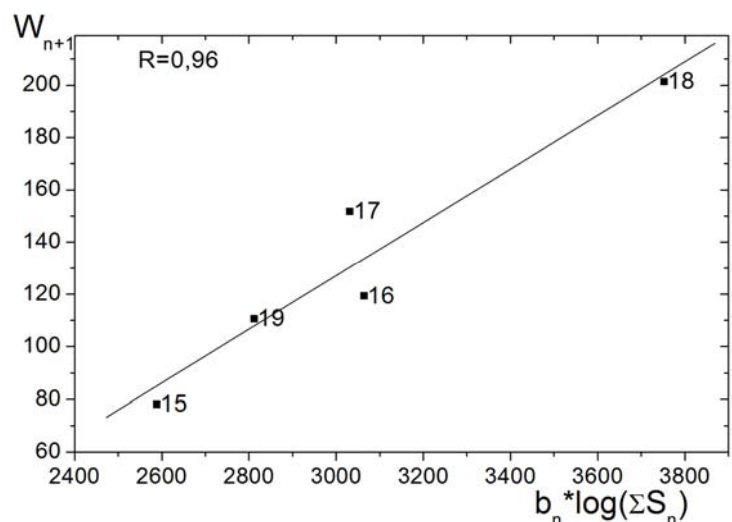

**Fig. 4.** Relation between the amplitude of the next activity cycle and the $b\log(\Sigma S\text{sp})$ index, compiled from the sum of all sunspot areas in a cycle and the magnetic field coupling coefficient in a cycle, where $b$ is the coefficient in the formula $B = a + b\log(S\text{sp})$.

## 4. LONG-TERM VARIATIONS IN THE G–O RULE

The G–O empirical rule (Vitinsky et al., 1986) was formulated for a pair of successive solar cycles. There are several definitions of this rule, but the main interpretation is as follows: the amplitude of an even activity cycle is smaller than the height of the next odd cycle. The Wolf number series, which was reconstructed by R. Wolf from 1748, is usually used to verify the G–O rule. However, as was indicated in (Hoyt and Schatten, 1998), this series has a rather large noise level since it was difficult to take into consideration small sunspots and several observations were not considered. Based on additional data, Hoyt and Schatten (1998) proposed a sunspot group index, reconstructed by them for 1610–1995.

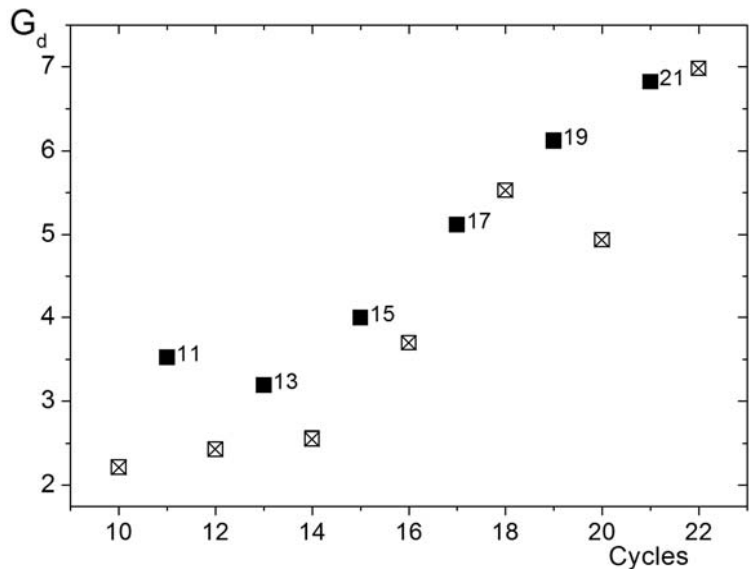

**Fig. 5.** Number of groups in a day during solar activity cycles according to the sunspot group index data. Odd and even cycles are marked with filled and open squares, respectively.

To characterize activity cycles, we can use the daily average group number in a cycle. Figure 5 presents such a number calculated based on the sunspot group index $G_d(k) = \sum_{T_k}^{T_{k+1}-1} Rg / Nd$ where $N_d$ is the number of observation days in cycle $k$ and $T_k$ is the time of the cycle $k$ onset. The times of the cycle onset and end we took from the NGDC site. During cycles 12–21, the average sunspot group number ($G_d$) in even cycles was smaller than in the next odd cycles, and this ratio is $G_d^{odd}/G_d^{even} \approx 1.39$. Therefore, the index of the daily average sunspot group number can be used to verify the G–O rule.

Figure 6 presents the $G_d^{odd}/G_d^{even}$ ratio from 1610 to 2009. The group number according to the (http://solarscience.msfc.nasa.gov) data for cycle 23 was added here to the daily data on the sunspot group number (Hoyt and Schatten, 1998). Figure 6 indicates that the $G_d^{odd}/G_d^{even}$ ratio corresponds to the standard definitions of the G–O rule after cycle №10 but also shows a smooth envelop for the cycles in the previous epoch, except several individual cycles. The line where this ratio is 1 was drawn for comparison. The data of cycle № –4/–3 are not presented.

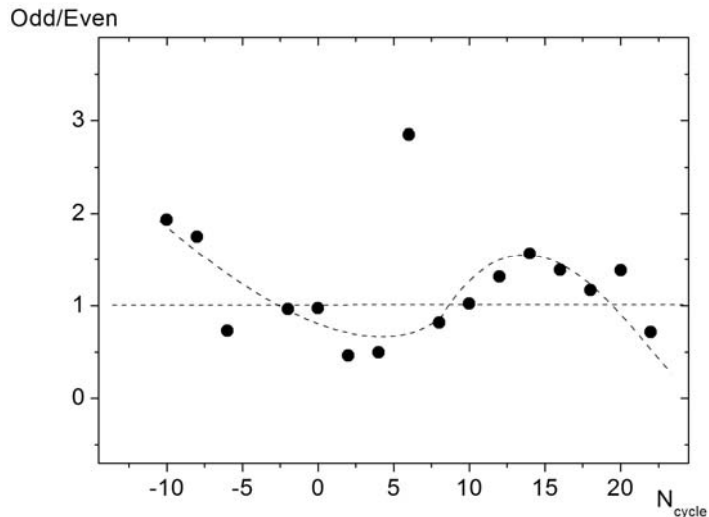

**Fig. 6.** Ratio of the daily average sunspot group number in an odd cycle to that in the previous even cycle.

## 5. CONCLUSIONS

An increase in activity in the middle of the 20th century was accompanied by a variation in the sunspot group properties. The sunspot number in groups increased during cycles 14–19. After cycle 19, this parameter tended to decrease (Figs. 2, 3). In the first half of the 20th century, the ratio of the umbrage area to the total group area varied (Fig. 3b) and the average magnetic field strength decreased (Fig. 3d) during the secular cycle activity growth stage. On the whole, this analysis confirms the assumption that the average sunspot number in a group varies following the secular cycle (Vitinsky et al., 1986). In this case, a decrease in the umbrage area with increasing area and sunspot number indicates that the magnetic flux is redistributed in sunspots from cycle to cycle. At the same time, different sunspot parameters have different durations of the secular cycles. The parameters related to the number of sunspots and sunspot groups have a duration of about ten cycles (Fig. 3a); the parameters related to the sunspot area, about eight activity cycles. It was also noted previously (Hathaway, 2010) that the secular cycle has different periods.

The relation between the sunspot area and the magnetic field changes during a secular cycle. In this case, the product of the total area logarithm into the coupling coefficient (*b*) characterizes the total sunspot magnetic field in the current cycle and is responsible for the next cycle level (Fig. 4). This fact can be the key to understanding the solar cyclicity.

Studying the G–O rule can also give important information regarding the solar cyclicity nature, specifically, the possible relic field, to which this effect is usually related (Mursula et al., 2001). Some authors consider that even cycles are constantly less intense than odd ones (Mursula et al., 2001; Nagovitsyn et al., 2009) and, therefore, introduce additional activity cycles. However, the pair of cycles 22 and 23 demonstrates that this rule is violated. Therefore, this rule possibly also reversed in the previous centuries and varies cyclically. The usage of the average group number in a cycle makes it possible to interpret the G–O rule as a long-term variation (Fig. 6). Long-term activity variations are possibly caused by a residual slowly varying (permanent) poloidal solar magnetic field. This field can nevertheless reverse its sign (which results in a reversal in the series of 22_year cycles) and modulate secular activity cycles. Such a permanent field is caused by "magnetic memory" below the sunspot generating zone (Tlatov, 2007). If this is the case, the reversal of the G–O rule in cycles 22–23 indicates that the reversal will also be valid for the next pair of even–odd cycles.

**ACKNOWLEDGMENTS**
This work was partially supported by the Russian Academy of Sciences and the Russian Foundation for Basic Research.

**REFERENCES**
- Hathaway, D.H., The solar cycle, *Living Rev. Solar Phys.*, 2010, vol. 7, no. 1.
- Hoyt, D.V. and Schatten, H., Group sunspot numbers: A new solar activity reconstruction, *Solar Phys.*, 1998, vol. 181, pp. 491–512.
- Lefe'vre, L. and Clette, F., A global small sunspot deficit at the base of the index anomalies of solar cycle 23, *Astron. Astrophys.*, 2011, vol. 536, p. L11.
- Mursula, K., Usoskin, I.G., and Kovaltsov, G.A., Persistent 22-year cycle in sunspot activity: Evidence for a relic solar magnetic field, *Solar Phys.*, 2001, vol. 198, pp. 51–56.
- Nagovitsyn, Yu.A., Nagovitsyna, E.Yu., and Makarova, V.V., The Gnevyshev–Ohl rule for physical parameters of the solar magnetic field: The 400-year interval, *Astron. Lett.,* 2009, vol. 35, pp. 564–571.
- Pevtsov, A. A., Bertello, L., Tlatov, A. G., Kilcik, A., Nagovitsyn Yu.A., and Cliver, E.W., Cyclic and long-term variation of sunspot magnetic fields, *Solar Phys.*, 2013 (in press).
- Tlatov, A.G., Proc. 11th Pulkovo Conference "Solar Activity Physical Nature and Prediction of Its Geophysical Manifestations", 2007, pp. 343–347.
- Vitinsky, Yu.I., Kopetsky, M., and Kuklin, G.V., *Statistika pyatnoobrazovatel'noi deyatel'nosti Solntsa* (Sunspot Formation Statistics), Moscow: Nauka, 1986.